\documentclass[%
reprint,
showpacs,preprintnumbers,
amsmath,amssymb,
aps,
prb,
]{revtex4-1}

\usepackage{graphicx}
\usepackage{dcolumn}
\usepackage{bm}
\usepackage{colortbl}

\begin{document}


\title{Quenched dynamics of entangled states in correlated quantum dots}

\author{N.\,S.\,Maslova$^{1}$}
\altaffiliation{}
\author{P.\,I.\,Arseyev$^{2}$}
\author{V.\,N.\,Mantsevich$^{1}$}
\altaffiliation{} \email{vmantsev@gmail.com}

\affiliation{%
$^{1}$Moscow State University, 119991 Moscow, Russia, $^{2}$ P.N.
Lebedev Physical Institute RAS, 119991 Moscow, Russia
}%

\date{\today }
\begin{abstract}
Time evolution of initially prepared entangled state in the system
of coupled quantum dots has been analyzed by means of two different
theoretical approaches: equations of motion for the all orders
localized electron correlation functions, considering interference
effects, and kinetic equations for the pseudo-particle occupation
numbers with constraint on the possible physical states. Results
obtained by means of different approaches were carefully analyzed
and compared with each other. Revealed direct link between
concurrence (degree of entanglement) and quantum dots pair
correlation functions allowed us to follow the changes of
entanglement during time evolution of the coupled quantum dots
system. It was demonstrated that the degree of entanglement can be
controllably tuned during the time evolution of quantum dots system.
\end{abstract}

 \keywords{} \maketitle

\section{Introduction}
One of the most interesting problems in the present-day nanophysics
is the controllable formation of entangled electronic states for use
in quantum information processing and cryptography. Coupled quantum
dots (QDs) systems recently seem to be promising candidates for
quantum information applications as single and two-electronic states
can be well initialized, processed and readout in such ultra-small
structures
\cite{Loss,Imamoglu,Yao,Blaauboer,Robledo,Nowack,Schulman,Maslova}.

Properties of entangled states are usually analyzed in the
stationary case. However, time evolution of spin and charge
configurations, initially prepared in coupled QDs, is also of great
interest as non-stationary characteristics could reveal new
information about the physical properties of nanoscale systems in
addition to the stationary ones
\cite{Bar-Joseph,Gurvitz_1,Arseyev_1,Mantsevich,Arseyev_2,Stafford_1,Hazelzet,Cota}.
Kinetics of initially prepared charge and spin states in quantum
dots systems is strongly governed by the high order localized
electrons correlation functions due to the presence of Coulomb
interaction \cite{Arseyev_3} and is also influenced by the
interference effects between electrons traveling through different
paths \cite{Wiel,Okazaki,Amasha}.

One of the challenges in the area of non-stationary electron
transport through coupled QDs is to prepare interacting few-level
systems with different initial states
\cite{Bayer,Creatore,Tsukanov,Yokoshi} - from simple product states
to complex entanglements. Various ideas for entangling of spatially
separated electrons were proposed, such as, by splitting Cooper
pairs \cite{Burkard} or by spin manipulation in QDs
\cite{Sanchez},\cite{Ciccarello}. In double correlated QDs entangled
state can appear as an eigenstate with particular number of
electrons \cite{Blaauboer},\cite{Burkard_1} or by sending an
electrical current through the nano-scale structure \cite{Busser}.
There are a lot of possible applications of entangled states in
nano-electronics \cite{Dowling}, including quantum information
processing \cite{Nielsen}. Most of the proposed schemes for quantum
computation deal with the spin control due to the localized spins
long decoherence times \cite{Hanson}. As it was recently shown,
entangled states in correlated QDs could reveal long relaxation
times due to the particular symmetry of investigated system.
Moreover, entangled states in correlated quantum dots can be well
controlled by applied bias voltage changing
\cite{Murgida},\cite{Kataoka} or by external laser pulses
\cite{Putaja},\cite{Saelen}.

Recently the potential of quantum information processing and quantum
computation results in numerous proposals of specific material
systems for creation and manipulation of entanglement in solid
state. System based on the coupled quantum dots with Coulomb
correlations has several appealing features: 1) single spin is a
natural qubit, 2) presence of strong Coulomb interaction within the
system creates entanglement even in the most easily experimentally
obtained ground state, 3) entangled quantum states in the coupled
QDs can be experimentally realized without such restrictions as for
two-impurity Kondo model. Moreover, the degree of entanglement can
change during the relaxation of initially prepared charge state in
double QD coupled to reservoir \cite{Maslova}.

In the present paper we analyze time evolution of initially prepared
entangled state in the correlated double QD due to the interaction
with an electronic reservoir. Two different approaches were
considered: the first one is based on the equations of motion for
all orders localized electron correlation functions and the second
one deals with the kinetic equations for pseudo-particle occupation
numbers, considering constraint on the possible physical states.
Results, obtained by means of these approaches were carefully
analyzed and compared with each other. It was demonstrated that both
approaches allow one to follow the changes of the system
entanglement during time evolution due to the direct link between
concurrence and quantum dots pair correlation functions. For
different initial mixed states entanglement could reveal
non-monotonic behavior and even increase considerably during the
relaxation processes in coupled quantum dots in the particular time
interval. So, one can tune the degree of entanglement during the
time evolution of correlated QDs. Proposed system is a good
candidate for quantum information protocol (QIP) realization with
the help of scanning tunneling microscopy/spectroscopy technique.

\section{Theoretical model}

We consider a system of two coupled correlated QDs connected to an
electronic reservoir. The Hamiltonian $\hat{H_{D}}$, describing
interacting quantum dots reads

\begin{eqnarray}
\hat{H}_{D}&=&\sum_{l=1,2,\sigma}\varepsilon_{l}c^{+}_{l\sigma}c_{l\sigma}+\sum_{l=1,2}U_{l}n_{ll}^{\sigma}n_{ll}^{-\sigma}+\nonumber\\&+&\sum_{\sigma}T(c_{1\sigma}^{+}c_{2\sigma}+c_{1\sigma}c_{2\sigma}^{+}),
\end{eqnarray}

where $\varepsilon_l$ ($l=1,2$) are the spin-degenerate
single-electron energy levels and $U_{l}$ is the on-site Coulomb
repulsion for the quantum dots double occupation.
Creation/annihilation of an electron with spin $\sigma=\pm1$ within
the dot is denoted by operators $c^{+}_{l\sigma}/c_{l\sigma}$ and
$n_{ll}^{\sigma}$ is the corresponding occupation number operator.
Coupling between the dots is described by tunneling transfer
amplitude $T$ which is considered to be independent on momentum and
spin.

Reservoir is modeled by the Hamiltonian:

\begin{eqnarray}
\hat{H}_{res}=\sum_{k\sigma}\varepsilon_{k}c^{+}_{k\sigma}c_{k\sigma},
\end{eqnarray}

where operator $c^{+}_{k\sigma}/c_{k\sigma}$ creates/annihilates an
electron with spin $\sigma$ and momentum $k$ in the lead. Coupling
between both dots and reservoir is described by the Hamiltonian:

\begin{eqnarray}
\hat{H}_{tun}=\sum_{k\sigma}t(c_{k\sigma}^{+}c_{l\sigma}+c_{l\sigma}^{+}c_{k\sigma}).
\end{eqnarray}

Tunneling amplitude $t$ is independent on momentum and spin. When
coupling between QDs exceeds the value of interaction with the
reservoir, one can use the basis of exact eigenfunctions and
eigenvalues of coupled QDs without interaction with the leads. In
this case all energies of single- and multi-electron states are well
known.

Two single electron states are present in the system and can be
described by the wave function

\begin{eqnarray}
\Psi_{i}^{\sigma}=\mu_{i}\cdot|0\uparrow\rangle|00\rangle+\nu_{i}\cdot|00\rangle|0\uparrow\rangle,
\end{eqnarray}

where basis functions $|0\uparrow\rangle|00\rangle$ and
$|00\rangle|0\uparrow\rangle$ describe the existence of single
electron with a given spin in each quantum dot. Single electron
energies

\begin{eqnarray}
\varepsilon_{a(s)}=\frac{\varepsilon_1+\varepsilon_2}{2}\pm\sqrt{\frac{(\varepsilon_1-\varepsilon_2)^{2}}{4}+T^{2}}
\end{eqnarray}

and coefficients $\mu_{i}$ and $\nu_{i}$ are determined by the
eigenstates of matrix:

\begin{eqnarray}
\begin{pmatrix}\varepsilon_{1} && -T\\
-T && \varepsilon_{2}\end{pmatrix}. \label{m1}\end{eqnarray}

Six two electron states exist in the system: two states with the
same electrons spin in each dot are given by the wave functions
$T^{+}=|\uparrow0\rangle|\uparrow0\rangle$ and
$T^{-}=|\downarrow0\rangle|\downarrow0\rangle$. Such states can be
formed only by electrons localized in the different dots. Four
states with the opposite spins can be described by the wave function

\begin{eqnarray}
\Psi_{j}^{\sigma-\sigma}&=&\alpha_{j}\cdot|\uparrow\downarrow\rangle|00\rangle+\beta_{j}\cdot|\downarrow0\rangle|0\uparrow\rangle+\nonumber\\&+&
\gamma_{j}\cdot|0\uparrow\rangle|\downarrow0\rangle+\delta_{j}\cdot|00\rangle|\uparrow\downarrow\rangle,\label{eq11}\nonumber\\
\end{eqnarray}

where basis wave functions $|\uparrow\downarrow\rangle|00\rangle$;
$|00\rangle|\uparrow\downarrow\rangle$ correspond to electrons
localized in the same quantum dot - the first one or the second one
and functions $|\downarrow0\rangle|0\uparrow\rangle$;
$|0\uparrow\rangle|\downarrow0\rangle$ describe the situation when
electrons are localized in different dots. Two electron states
energies and coefficients $\alpha_{j}$, $\beta_{j}$, $\gamma_{j}$,
$\delta_{j}$ are determined by the eigenvalues and eigenvectors of
matrix:

\begin{eqnarray}
\begin{pmatrix}2\varepsilon_{1}+U_{1} && -T && -T && 0 \\
-T && \varepsilon_{1}+\varepsilon_{2} && 0 && -T\\
-T && 0 && \varepsilon_{1}+\varepsilon_{2} && 0\\
0 && -T && -T && 2\varepsilon_{2}+U_{2}\end{pmatrix}.
\label{m2}\end{eqnarray}

These are low energy singlet $S^{0}$ and triplet $T^{0}$ states and
excited singlet ($S^{0*}$) and triplet states ($T^{0*}$). Low energy
triplet state $T^{0}$ with energy $\varepsilon_1+\varepsilon_2$
exists for any values of QDs energy levels $\varepsilon_l$ and
Coulomb interaction $U_l$. Corresponding coefficients in
Eq.(\ref{eq11}) are $\alpha=\delta=0$ and
$\beta=-\gamma=\frac{1}{\sqrt{2}}$.

Two three electron states with the wave function

\begin{eqnarray}
\Psi_{m}^{\sigma\sigma-\sigma}&=&p_{m}|\uparrow\downarrow\rangle|\uparrow0\rangle+q_{m}|\uparrow0\rangle|\uparrow\downarrow\rangle\nonumber\\
m&=&\pm1
\end{eqnarray}

are present in the system. In this case basis functions
$|\uparrow\downarrow\rangle|\uparrow0\rangle$ and
$|\uparrow0\rangle|\uparrow\downarrow\rangle$ describe the
situation, when one of the dots is fully occupied by two electrons
with opposite spins and only single electron with a given spin is
present in another dot. Coefficients $p_{m}$, $q_{m}$ and energies
are determined by the eigenvectors and eigenvalues of matrix:

\begin{eqnarray}
\begin{pmatrix}2\varepsilon_{1}+\varepsilon_{2}+U_{1} && -T\\
-T && 2\varepsilon_{2}+\varepsilon_{1}+U_{2}\end{pmatrix}.
\label{m3}\end{eqnarray}

Finally, single four-electron state exists in the system with the
wave function

\begin{eqnarray}
\Psi_{n}=|\uparrow\downarrow\rangle|\uparrow\downarrow\rangle.
\end{eqnarray}

In this case both quantum dots are fully occupied.

\subsection{Equations of motion for localized electron correlation functions}

Coupling to reservoir leads to the changing of the dots occupation
due to the tunneling processes. We now derive kinetic equations for
bilinear combinations of the Heisenberg operators
$c_{l\sigma}^{+}/c_{l\sigma}$, which allow to analyze the dynamics
of localized electron occupation numbers and high order correlation
functions due to the coupling to reservoir:

\begin{eqnarray}
c_{1\sigma}^{+}c_{1\sigma}=\hat n_{11}^{\sigma}(t);\quad
c_{2\sigma}^{+}c_{2\sigma}=\hat n_{22}^{\sigma}(t);\nonumber\\
c_{1\sigma}^{+}c_{2\sigma}=\hat n_{12}^{\sigma}(t);\quad
c_{2\sigma}^{+}c_{1\sigma}=\hat n_{21}^{\sigma}(t).
\end{eqnarray}

We consider time evolution of initially prepared state in the case
of "empty" reservoir in a wide band limit approximation and for deep
energy levels ($\frac{|\varepsilon_i-\varepsilon_F|}{\Gamma}>>1$,
where $\Gamma=\pi\nu_0t^{2}$, $\nu_0$ - is unperturbated density of
states in the reservoir) when applied bias is equal to zero.

By means of Heisenberg equations of motion one can get closed system
of equations for localized electrons occupation numbers exactly
taking into account correlations of all orders
\cite{Mantsevich},\cite{Arseyev_2} (for weak tunneling coupling
between QDs and reservoir). Kinetic equations describe time
evolution of the electron occupation numbers in the proposed system:

\begin{eqnarray}
\frac{\partial}{\partial
t}\hat n_{11}^{\sigma}&=&-\Gamma (\hat n_{21}^{\sigma}+\hat n_{12}^{\sigma})+iT(\hat n_{21}^{\sigma}-\hat n_{12}^{\sigma})-2\Gamma\hat n_{11}^{\sigma},\nonumber\\
\frac{\partial}{\partial
t}\hat n_{22}^{\sigma}&=&-\Gamma (\hat n_{21}^{\sigma}+\hat n_{12}^{\sigma})-iT(\hat n_{21}^{\sigma}-\hat n_{12}^{\sigma})-2\Gamma\hat n_{22}^{\sigma},\nonumber\\
\frac{\partial}{\partial t}\hat n_{21}^{\sigma}&=&-\Gamma(\hat
n_{11}^{\sigma}+\hat n_{22}^{\sigma})+iT(\hat n_{11}^{\sigma}-\hat
n_{22}^{\sigma})-i(\xi-2i\Gamma)\hat n_{21}^{\sigma} -\nonumber\\&-&
iU_{11}n_{21}^{\sigma}n_{11}^{-\sigma}
+iU_{22} n_{21}^{\sigma}n_{22}^{-\sigma},\nonumber\\
\frac{\partial}{\partial t}\hat n_{12}^{\sigma}&=&-\Gamma(\hat
n_{11}^{\sigma}+\hat n_{22}^{\sigma})+iT(\hat n_{11}^{\sigma}-\hat
n_{22}^{\sigma}) +i(\xi+2i\Gamma)\hat n_{12}^{\sigma}
+\nonumber\\&+& iU_{11}n_{12}^{\sigma}n_{11}^{-\sigma} -iU_{22} \hat
n_{12}^{\sigma}n_{22}^{-\sigma},\ \label{system}
\end{eqnarray}

where $\xi=\varepsilon_1-\varepsilon_2$ is the detuning between
energy levels in the dots. The first term in each right-hand part of
Eqs.(\ref{system}) [$\Gamma\cdot(\hat n_{21}^{\sigma}+\hat
n_{12}^{\sigma})$ or $\Gamma\cdot(\hat n_{11}^{\sigma}+\hat
n_{22}^{\sigma})$] appears due to the interference effects caused by
the charge relaxation to reservoir through different possible
channels, similar to the Fano effect. These terms are absent if only
one quantum dot is coupled to reservoir. System of
Eqs.(\ref{system}) contains pair correlation operators
$\widehat{K}^{\sigma\sigma^{'}}_{lrl^{'}r^{'}}=\widehat{n}^{\sigma}_{lr}\widehat{n}^{\sigma^{'}}_{l^{'}r^{'}}$,
which also determine relaxation and, consequently, should be
calculated. If one is interested in relaxation dynamics of the
two-electron initial state, only pair correlation functions should
be retained as the situation of "empty" reservoir is considered.

Let us introduce the correlations operators
$\widehat{K}^{\sigma\sigma^{'}}_{lrl^{'}r^{'}}$ averaged values
$K^{\sigma\sigma^{'}}_{lrl^{'}r^{'}}=<c_{l\sigma}^{+}c_{r\sigma}c_{l^{'}\sigma^{'}}^{+}c_{r^{'}\sigma^{'}}>$
- elements of $\widehat{\textbf{K}}$ $4\times4$ matrix. System of
equations for the pair correlation functions can be written in the
compact matrix form (symbol $[\quad]$ means commutation and symbol
$\{\quad\}$ - anti-commutation)

\begin{eqnarray}
i\frac{\partial}{\partial
t}\widehat{\textbf{K}}=[\widehat{\textbf{K}},\widehat{H}^{'}]+\{\widehat{\textbf{K}},\widehat{\Gamma}\},
\label{system_compact}
\end{eqnarray}

where matrix $\widehat{H}^{'}$ has the following form

\begin{eqnarray}
\widehat{H}^{'}= \left(\begin{array}{ccccc}
0 & T+i\Gamma & T-i\Gamma & 0\\
T-i\Gamma & \xi+U_{1} & 0 & T-i\Gamma\\
T+i\Gamma & 0 & -\xi+U_{2} & T+i\Gamma\\
0 & T+i\Gamma & T-i\Gamma & 0\\
\end{array}\right)
\end{eqnarray}

and $\widehat{\Gamma}$ is the relaxation diagonal $4\times4$ matrix
with non-zero elements $\Gamma_{nn}=-2i\Gamma$.

System of equations (\ref{system})-(\ref{system_compact}) for the
two-electron pure state $|\Psi_{j}^{\sigma-\sigma}\rangle$ time
evolution can be solved with the following initial conditions:
$n_{11}^{\sigma}(0)=\alpha^{2}+\beta^{2}$;
$n_{12}^{\sigma}(0)=n_{21}^{\sigma}(0)=\alpha\gamma+\beta\delta$;
$n_{22}^{\sigma}(0)=\delta^{2}+\gamma^{2}$;
$K_{1111}^{\sigma-\sigma}=\alpha^{2}$;
$K_{2222}^{\sigma-\sigma}=\delta^{2}$;
$K_{1122}^{\sigma-\sigma}=\beta^{2}$;
$K_{2211}^{\sigma-\sigma}=\gamma^{2}$;
$K_{1221}^{\sigma-\sigma}=K_{2112}^{\sigma-\sigma}=\beta\gamma$;
$K_{2121}^{\sigma-\sigma}=K_{1212}^{\sigma-\sigma}=\alpha\delta$;
$K_{1211}^{\sigma-\sigma}=K_{2111}^{\sigma-\sigma}=\gamma\alpha$;
$K_{1112}^{\sigma-\sigma}=K_{1121}^{\sigma-\sigma}=\alpha\beta$;
$K_{1222}^{\sigma-\sigma}=K_{2122}^{\sigma-\sigma}=\beta\delta$;
$K_{2221}^{\sigma-\sigma}=K_{2212}^{\sigma-\sigma}=\gamma\delta$,
where coefficients $\alpha$, $\beta$, $\gamma$ and $\delta$ are
given by the eigenvectors of matrix (\ref{m2}).

For initial mixed two-electron state with density matrix
$\rho(0)=\sum_{j,\sigma,\sigma'}N_{j}^{\sigma\sigma'}(0)|\Psi_{j}^{\sigma\sigma'}\rangle\langle\Psi_{j}^{\sigma\sigma'}|$,
where $N_{j}^{\sigma\sigma'}(0)$
($j=S^{0},T^{0},S^{0*},T^{0*},T^{\pm}$) is the occupation number of
$j$ two-electron state at $t=0$, initial conditions for second order
correlation functions and for first order correlators are

\begin{eqnarray}
K_{lrl'r'}^{\sigma-\sigma}(0)=Sp[\widehat{\rho}(0)\widehat{K}_{lrl'r'}^{\sigma-\sigma}]
\end{eqnarray}

and

\begin{eqnarray}
n_{lr}^{\sigma}(0)=Sp[\widehat{\rho}(0)\widehat{n}_{lr}^{\sigma}].
\end{eqnarray}

We'll consider time evolution of singlet $S^{0}$ and triplet $T^{0}$
initial states because excited $S^{0*}$ and $T^{0*}$ states are
separated by Coulomb gap. One can also exclude states $T^{\pm}$ at
low temperature by introducing weak exchange interaction with
exchange constant $J_{z}>0$:

\begin{eqnarray}
\hat{H}_{ex}=J_{z}\cdot(n_{11}^{\sigma}-n_{11}^{-\sigma})\cdot(n_{22}^{\sigma}-n_{22}^{-\sigma}).
\end{eqnarray}

Consequently, initial two-electron density matrix can be written as:

\begin{eqnarray}
\rho(0)=N_{S^{0}}(0)|S^{0}\rangle\langle
S^{0}|+N_{T^{0}}(0)|T^{0}\rangle\langle T^{0}|.\label{initial}
\end{eqnarray}

For singlet initial state $S^{0}$ coefficients $\alpha$, $\beta$,
$\gamma$ and $\delta$ are determined as an eigenvector of matrix
(\ref{m2}) corresponding to its minimal eigenvalue, $N_{S^{0}}(0)=1$
and $N_{T^{0}}(0)=0$. For the triplet initial state $T^{0}$
coefficients $\alpha=\delta=0$ and
$\beta=-\gamma=\frac{1}{\sqrt{2}}$, $N_{S^{0}}(0)=0$ and
$N_{T^{0}}(0)=1$.

\subsection{Entangled states in correlated quantum dots}

Electron states in the correlated quantum dots can be entangled.
Entangled state is characterized by non-zero value of concurrence
$C$ \cite{Wootters}. Concurrence for pure state
$|\Psi_{j}^{\sigma\sigma'}\rangle$ is determined as
$C=|\langle\Psi_{j}^{\sigma\sigma'}|\widetilde{\Psi}_{j}^{\sigma\sigma'}\rangle|$,
where $|\widetilde{\Psi}_{j}^{\sigma\sigma'}\rangle$ is the "spin
flipped" state $|\Psi_{j}^{\sigma\sigma'}\rangle$. For mixed state
concurrence $C=max\{0,\lambda_1-\sum_{i}\lambda_i\}$, where
$\{\lambda_i\}$ are square roots of matrix $\widetilde{\rho}\rho$
($\widetilde{\rho}$ is the "spin flipped" matrix $\rho$) eigenvalues
arranged in the decreasing order. For the initial two-electron
entangled pure state $|\Psi_{j}^{\sigma\sigma'}\rangle$ with
opposite spins \cite{Maslova}:

\begin{eqnarray}
C=|\alpha^{2}+\delta^{2}+2\cdot\beta\cdot\gamma|.
\label{con}\end{eqnarray}

During time evolution of initial state system entanglement changes.
To follow these changes concurrence could be expressed through the
time dependent correlation functions. We'll demonstrate that for
arbitrary mixed state of two correlated quantum dots the concurrence
$C$ can be determined through the mean value of particular
combination of pair correlation functions
$\widehat{K}_{lrl^{'}r^{'}}^{\sigma-\sigma}$:

\begin{eqnarray}
C=\langle
\widehat{K}_{1111}^{\sigma-\sigma}+\widehat{K}_{1221}^{\sigma-\sigma}+\widehat{K}_{2112}^{\sigma-\sigma}+\widehat{K}_{2222}^{\sigma-\sigma}\rangle.\label{eq00}
\end{eqnarray}

Let us introduce operator $\widehat{K}^{'}$, which can be written as
a combination of pair correlation functions operators:
\begin{eqnarray}
\widehat{K}^{'}=\widehat{K}_{1111}^{\sigma-\sigma}+\widehat{K}_{1221}^{\sigma-\sigma}+\widehat{K}_{2112}^{\sigma-\sigma}+\widehat{K}_{2222}^{\sigma-\sigma}.
\end{eqnarray}

Acting by the operator $\widehat{K}^{'}$ on the wave function
$|\Psi_{j}^{\sigma\sigma'}\rangle$ one obtains "spin flipped" wave
function $|\widetilde{\Psi}_{j}^{\sigma\sigma'}\rangle$:

\begin{eqnarray}
\widehat{K}^{'}|\Psi_{j}^{\sigma\sigma'}\rangle=|\widetilde{\Psi}_{j}^{\sigma\sigma'}\rangle.
\end{eqnarray}

For any wave function $|\Psi_{j}^{\sigma\sigma'}\rangle$:

\begin{eqnarray}
|\langle\Psi_{j}^{\sigma\sigma'}|\widehat{K}^{'}|\Psi_{j}^{\sigma\sigma'}\rangle|=|\langle\Psi_{j}^{\sigma\sigma'}|\widetilde{\Psi}_{j}^{\sigma\sigma'}\rangle|=C.
\end{eqnarray}

If $\{|\Psi_{j}^{\sigma\sigma'}\rangle\}$ are the two-electron
eigenfunctions of the Hamiltonian $\widehat{H}$, two particle
density matrix can be written as
$\rho=\sum_{j}|\Psi_{j}^{\sigma\sigma'}\rangle\langle\Psi_{j}^{\sigma\sigma'}|N_{j}^{\sigma\sigma'}$.
For simplicity we'll further omit spin indexes in
$|\Psi_{j}^{\sigma\sigma'}\rangle$.  The following relations take
place: $|\langle
\Psi_{j}|\widehat{K}^{'}|\widetilde{\Psi}_{i}\rangle|=\delta_{ij}$
and $|\langle
\Psi_{i^{'}}|\widehat{K}^{'2}|\Psi_{i}\rangle|=\delta_{ii^{'}}=\sum_{j}\langle
\Psi_{i^{'}}|\widehat{K}^{'}|\widetilde{\Psi}_{j}\rangle\langle\widetilde{\Psi}_{j}|\widehat{K}^{'}|\Psi_{i}\rangle$.

Let us prove that
\begin{eqnarray}
\langle \Psi_{j}|\widetilde{\rho}\rho|\Psi_{i}\rangle=\langle
\Psi_{j}|\widehat{K}^{'}\rho\widehat{K}^{'}\rho|\Psi_{i}\rangle.
\label{eq0}\end{eqnarray}

Really:

\begin{eqnarray}
\langle \Psi_{j}|(\widehat{K}^{'}\rho)^{2}|\Psi_{i}\rangle&=&
\sum_{i_{1}}\langle
\Psi_{j}|\widehat{K}^{'}|\Psi_{i_{1}}\rangle\langle
\Psi_{i_{1}}|\widehat{K}^{'}|\Psi_{i}\rangle
N_{i}N_{i_{1}}=\nonumber\\&=&\sum_{i_{1}}\langle
\Psi_{j}|\widetilde{\Psi}_{i_{1}}\rangle\langle
\Psi_{i_{1}}|\widetilde{\Psi}_{i}\rangle N_{i}N_{i_{1}}
\label{eq1}\end{eqnarray}

and

\begin{eqnarray}
\langle
\Psi_{j}|\widetilde{\rho}\rho|\Psi_{i}\rangle&=&\sum_{i_{1}}N_{i}N_{i_{1}}\langle
\Psi_{j}|\widetilde{\Psi}_{i_{1}}\rangle\langle\widetilde{\Psi}_{i_{1}}|\Psi_{i}\rangle=\nonumber\\&=&
\delta_{ij}\sum_{i_{1}}N_{i}N_{i_{1}}|\langle
\Psi_{j}|\widetilde{\Psi}_{i_{1}}\rangle|^{2}.\nonumber\\
\label{eq2}\end{eqnarray}

Comparing expressions (\ref{eq1}) and (\ref{eq2}), one can find that
statement (\ref{eq0}) is valid and matrixes
$\|\widetilde{\rho}\rho\|_{ji}$ and
$\|(\widehat{K}^{'}\rho)^{2}\|_{ji^{'}}$ have the same eigenvalues.
If $\widetilde{\lambda}_{p}$ are the eigenvalues of matrix
$\|\widetilde{\rho}\rho\|_{ji}$ and $\lambda_{i}$ are the
eigenvalues of matrix $\|\widehat{K}^{'}\rho\|_{ji^{'}}$, then
$\lambda^{2}_{i}=\widetilde{\lambda}_{p}$. So,

\begin{eqnarray}
\langle\widehat{K}^{'}\rangle=Tr(\widehat{K}^{'}\rho)&=&\sum_{ij}\langle
j|\widehat{K}^{'}|i\rangle\langle i|j\rangle
\lambda_i=\sum_{i}\lambda_i\langle
i|\widetilde{i}\rangle.\nonumber\\
\end{eqnarray}

Relative sign of $\lambda_{i}=\pm\sqrt{\widetilde{\lambda}_{p}}$ is
determined by the sign of $\langle i|\widetilde{i}\rangle=\pm1$ for
singlet and triplet state.

Moreover, for the eigenstates:

\begin{eqnarray}
|\langle S^{0}|\widetilde{S^{0}}\rangle|=|\langle
T^{0}|\widetilde{T^{0}}\rangle|=|\langle
T^{\mp}|\widetilde{T}^{\pm}\rangle|=1.
\end{eqnarray}

Thus $\langle\widehat{K}^{'}\rangle$ can be expressed through
$\widetilde{\lambda}_{p}$, arranged in decreasing order:
$\langle\widehat{K}^{'}\rangle=(\sqrt{\widetilde{\lambda}_{1}}-\sum_{i>1}\sqrt{\widetilde{\lambda}_{i}})$.
So, the common definition of concurrence $C$ (see Ref.
\onlinecite{Wootters}) can be rewritten as

\begin{eqnarray}
C=max\{0,\langle\widehat{K}^{'}\rangle\}. \label{eq5}
\end{eqnarray}

\begin{figure}
\includegraphics[width=80mm]{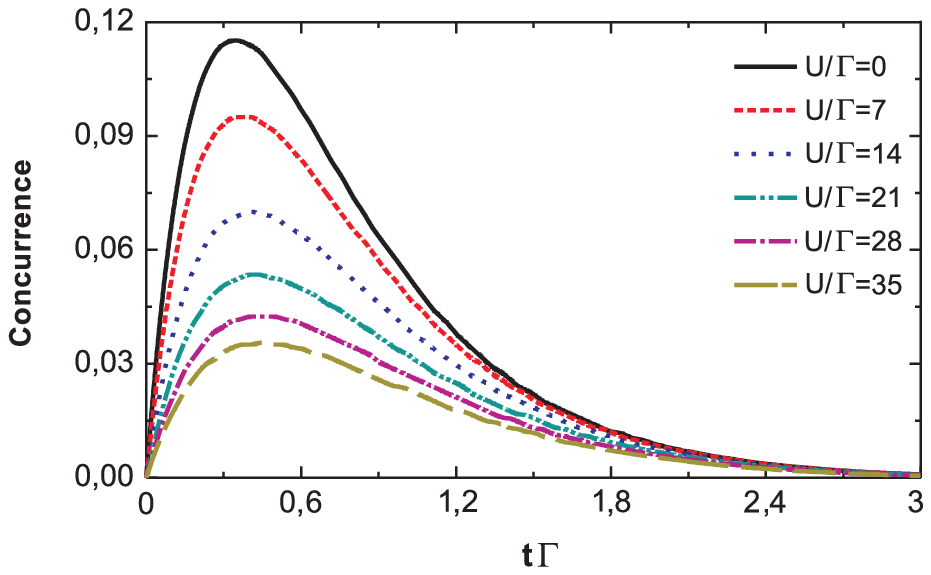}%
\caption{(Color online) Concurrence time evolution for different
values of Coulomb interaction $U_{1}/\Gamma=U_{2}/\Gamma=U/\Gamma$.
$\varepsilon_1/\Gamma=\varepsilon_1/\Gamma=7$, $T/\Gamma=2$, and
$\Gamma=1$. Initial conditions are: $N_{S}(0)=0.5$, $N_{T}(0)=0.5$.}
\label{figure1}
\end{figure}

\begin{figure}
\includegraphics[width=80mm]{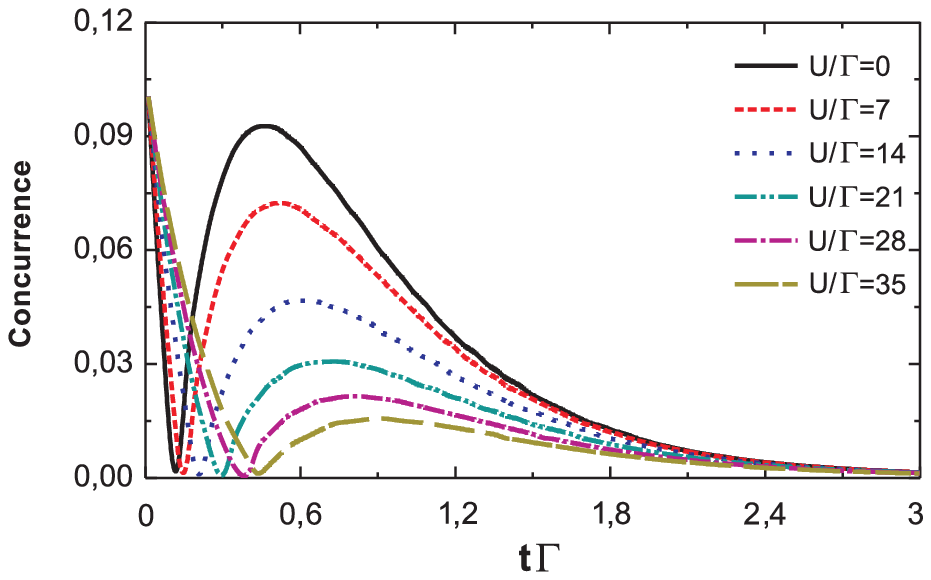}%
\caption{(Color online) Concurrence time evolution for different
values of Coulomb interaction $U_{1}/\Gamma=U_{2}/\Gamma=U/\Gamma$.
$\varepsilon_1/\Gamma=\varepsilon_1/\Gamma=7$, $T/\Gamma=2$, and
$\Gamma=1$. Initial conditions are: $N_{S}(0)=0.55$,
$N_{T}(0)=0.45$.} \label{figure2}
\end{figure}

\begin{figure}
\includegraphics[width=80mm]{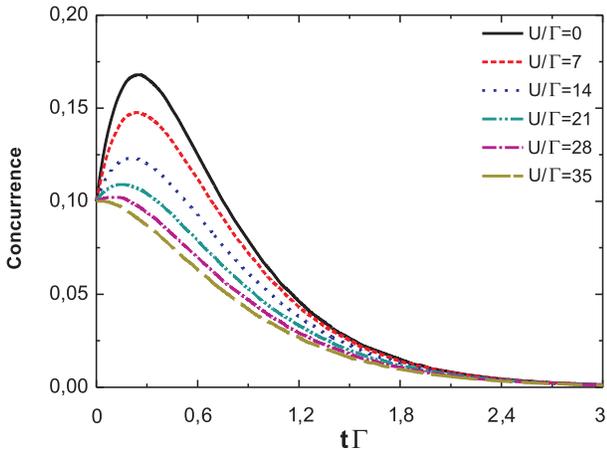}%
\caption{(Color online) Concurrence time evolution for different
values of Coulomb interaction $U_{1}/\Gamma=U_{2}/\Gamma=U/\Gamma$.
$\varepsilon_1/\Gamma=\varepsilon_1/\Gamma=7$, $T/\Gamma=2$, and
$\Gamma=1$. Initial conditions are: $N_{S}(0)=0.45$,
$N_{T}(0)=0.55$.} \label{figure3}
\end{figure}

The behavior of time dependent quantum-dot system concurrence
calculated by means of Eqs. (\ref{system}), (\ref{eq00}),
(\ref{eq5}) for different initial conditions is demonstrated in
Figs.\ref{figure1}-\ref{figure3}. Figs.\ref{figure1}, \ref{figure2}
demonstrate an important fact, that concurrence - the degree of
entanglement, can increase during the relaxation processes in the
system of coupled QDs, caused by the presence of on-site Coulomb
correlations and interaction with the reservoir. Results depicted in
Fig.\ref{figure3} reveal the possibility of system switching between
entangled and unentangled (concurrence is equal to zero) states
during the relaxation process.

\subsection{Kinetic equations in pseudo-particle formalism}

Another method of quantum-dot system dynamics analysis is based on
the pseudo-particles formalism \cite{Coleman}. Each pseudo-particle
corresponds to particular eigenstate of the system. Transitions
between the states with different number of electrons caused by
coupling to reservoir can be analyzed in terms of pseudo-particle
operators with constraint on the possible physical states (the
number of pseudo-particles). Consequently, electron operator
$c_{l\sigma}^{+}$ $(l=1,2)$ can be written in terms of
pseudo-particle operators \cite{Maslova}:

\begin{eqnarray}
c_{l\sigma}^{+}&=&\sum_{i}X_{i}^{\sigma
l}f_{i\sigma}^{+}b+\sum_{j,i}Y_{ji}^{\sigma-\sigma
l}d_{j}^{+\sigma-\sigma}f_{i-\sigma}+\\&+&\sum_{i}Y_{i}^{\sigma\sigma
l}d^{+\sigma\sigma}f_{i\sigma}+\sum_{m,j}Z_{mj}^{\sigma\sigma-\sigma
l}\psi_{m-\sigma}^{+}d_{j}^{\sigma-\sigma}+\nonumber\\&+&\sum_{m}Z_{m}^{\sigma-\sigma-\sigma
l}\psi_{m\sigma}^{+}d^{-\sigma-\sigma}+\sum_{m}W_{m}^{\sigma-\sigma-\sigma
l}\varphi^{+}\psi_{m\sigma}\nonumber\ \label{sb}\end{eqnarray}

with constraint on the possible physical states

\begin{eqnarray}
\widehat{n}_{b}+\sum_{i\sigma}\widehat{n}_{fi\sigma}+\sum_{j\sigma\sigma^{'}}\widehat{n}_{dj}^{\sigma\sigma^{'}}+\sum_{m\sigma}\widehat{n}_{\psi
m\sigma}+\widehat{n}_{\varphi}=1, \label{const}
\end{eqnarray}

where $f_{\sigma}^{+}(f_{\sigma})$ and
$\psi_{\sigma}^{+}(\psi_{\sigma})$ - are pseudo-fermion creation
(annihilation) operators for the electronic states with one and
three electrons correspondingly. $b^{+}(b)$,
$d_{\sigma}^{+}(d_{\sigma})$ and $\varphi^{+}(\varphi)$ - are slave
boson operators, which correspond to the states without any
electrons, with two electrons or four electrons. Operators
$\psi_{m-\sigma}^{+}$ - describe system configuration with two spin
up electrons $\sigma$ and one spin down electron $-\sigma$ in the
symmetric and asymmetric states.

Further we'll consider only single- and low energy double-occupied
states, because the excited double-occupied states, three- and four-
particle states are separated by the Coulomb gap. Consequently, all
the terms containing  $\varphi^{+}$  and $\psi_{m-\sigma}^{+}$ in
expression (\ref{sb}) are omitted. Matrix elements $X_{i}^{\sigma
l}$, $Y_{ji}^{\sigma-\sigma l}$ and $Y_{ji}^{\sigma\sigma l}$ can be
defined as:

\begin{eqnarray}
X_{i}^{\sigma l}&=&\langle\Psi_{i}^{\sigma}|c_{l\sigma}^{+}|0\rangle,\nonumber\\
Y_{ji}^{\sigma-\sigma l}&=&\langle\Psi_{j}^{\sigma-\sigma}|c_{l\sigma}^{+}|\Psi_{i}^{-\sigma}\rangle,\nonumber\\
Y_{ji}^{\sigma\sigma l}&=&\langle\Psi_{j}^{\sigma\sigma}|c_{l\sigma}^{+}|\Psi_{i}^{\sigma}\rangle.\nonumber\\
\end{eqnarray}

So, taking into account constraint on the possible physical states
the following non-stationary system of equations can be obtained for
the pseudo-particle occupation numbers $N_{i}^{\sigma}$,
$N_{j}^{\sigma-\sigma}$, $N_{j}^{\sigma\sigma}$ and $N_{b}$ by means
of Heisenberg equations:

\begin{eqnarray}
\frac{\partial N_{j}^{\sigma-\sigma}}{\partial
t}&=&-2\Gamma\sum_{i\sigma
}|Y_{ji}^{\sigma-\sigma}|^{2}\cdot N_{j}^{\sigma-\sigma},\nonumber\\
\frac{\partial N_{i}^{\sigma}}{\partial
t}&=&2\Gamma\sum_{j}|Y_{ji}^{\sigma-\sigma}|^{2}N_{j}^{\sigma-\sigma}-\nonumber\\&-&2\Gamma|X_{i}^{\sigma}|^{2}N_{i}^{\sigma}+2\Gamma\sum_{j
}|Y_{ji}^{\sigma\sigma}|^{2}\cdot N_{j}^{\sigma\sigma},\nonumber\\
\frac{\partial N_{b}}{\partial t}&=&2\Gamma\sum_{i\sigma
}|X_{i}^{\sigma}|^{2}\cdot N_{i}^{\sigma},\nonumber\\
\frac{\partial N_{j}^{\sigma\sigma}}{\partial
t}&=&-2\Gamma\sum_{i}|Y_{ji}^{\sigma\sigma}|^{2}\cdot
N_{j}^{\sigma\sigma}, \label{sys1}
\end{eqnarray}

where matrix elements can be easily expressed through the elements
of matrixes (\ref{m1}), (\ref{m2}) and (\ref{m3}) eigenvectors:

\begin{eqnarray}
|X_{i}^{\sigma}|^{2}&=&|\nu_i+\mu_i|^{2},\nonumber\\
|Y_{ji}^{\sigma-\sigma}|^{2}&=&|\alpha_j\mu_i+\beta_j\nu_i+\gamma_j\mu_i+\delta_j\nu_i|^{2},\nonumber\\
|Y_{ji}^{\sigma\sigma}|^{2}&=&|\nu_i+\mu_i|^{2}.
\end{eqnarray}

Depending on the tunneling barrier width and height typical
tunneling rate $\Gamma$ can vary from $10\mu eV$ \cite{Amaha} to
$1\div5$ meV\cite{Fransson}. These equations conserve the total
number of pseudo-particles:

\begin{eqnarray}
N_{b}+\sum_{i\sigma}N_{i}^{\sigma}+\sum_{j\sigma\sigma^{'}}N_{j}^{\sigma\sigma^{'}}=const.
\label{limit}
\end{eqnarray}

So, Eqs.(\ref{sys1}) provide the fulfilment of constraint

\begin{eqnarray}
N_{b}+\sum_{i\sigma}N_{i}^{\sigma}+\sum_{j\sigma\sigma^{'}}N_{j}^{\sigma\sigma^{'}}=1
\label{limit1}
\end{eqnarray}

during time evolution, if it occurs at the initial time moment.

System of Eqs.(\ref{sys1}) can be solved analytically with initial
conditions $N_{j}^{\sigma\sigma^{'}}(0)=N_{j}$,
$N_{a}^{\sigma}(0)=0$, $N_{s}^{\sigma}(0)=0$ and $N_{b}(0)=0$
($\sum_{j}N_{j}=1$). For initial generally mixed singlet-triplet
state (\ref{initial}):

\begin{eqnarray}
N_{j}^{\sigma-\sigma}(t)&=&N_{j}\cdot e^{-2\lambda_j t},\nonumber\\
N_{a}^{\sigma}(t)&=&\sum_{j}[\frac{\lambda_{ja}}{2\lambda_j-\lambda_a}\cdot(e^{-\lambda_at}-e^{-2\lambda_j
t})]\cdot N_{j},\nonumber\\
N_{s}^{\sigma}(t)&=&\sum_{j}[\frac{\lambda_{js}}{2\lambda_j-\lambda_s}\cdot(e^{-\lambda_st}-e^{-2\lambda_j
t})]\cdot N_j,\nonumber\\
N_{b}(t)&=&1-N_{dj}^{\sigma-\sigma}(t)-\sum_\sigma
N_{a}^{\sigma}(t)-\sum_\sigma N_{s}^{\sigma}(t),\label{ppsys}\nonumber\\
\end{eqnarray}

where

\begin{eqnarray}
\lambda_{a(s)}&=&2\Gamma\cdot|\mu_{a(s)}+\nu_{a(s)}|^{2},\nonumber\\
\lambda_{ja(s)}&=&2\Gamma\cdot|\alpha_j\mu_{a(s)}+\beta_j\nu_{a(s)}+\delta_j\nu_{a(s)}+\gamma_j\mu_{a(s)}|^{2},\nonumber\\
\lambda_j&=&\sum_{i=a,s}\lambda_{ji}.
\end{eqnarray}

Electron occupation numbers $N_{el}$ can be determined through the
pseudo-particle occupation numbers considering spin degrees of
freedom by the following expression:

\begin{eqnarray}
N_{el}(t)=\sum_{j}N_{j}\cdot[2e^{-2\lambda_j
t}+2\sum_{i=a,s}\frac{\lambda_{ij}}{2\lambda_j-\lambda_i}(e^{-\lambda_it}-e^{2\lambda_j
t})].\nonumber\\
\end{eqnarray}

According to concurrence definition through the eigenvalues of
matrix $\widetilde{\rho}\rho$ for initial state (\ref{initial}):

\begin{eqnarray}
C(t)=max(0,|N_{S^{0}}(t)-N_{T^{0}}(t)|),
\end{eqnarray}

where

\begin{eqnarray}
N_{S^{0}}(t)&=&N_{S^{0}}(0)\cdot e^{-2\lambda_{S^{0}}t},\nonumber\\
N_{T^{0}}(t)&=&N_{T^{0}}(0)\cdot e^{-2\lambda_{T^{0}}t},\nonumber\\
\lambda_{S^{0}}&=&|\alpha+\beta|^{2}\cdot(\lambda_s+\lambda_a),\nonumber\\
\lambda_{T^{0}}&=&\frac{1}{2}\cdot(\lambda_s+\lambda_a).
\end{eqnarray}

Concurrence time evolution for different initial conditions and
values of Coulomb correlations is shown in
Figs.\ref{figure1}-\ref{figure3}. Results obtained by both
approaches exactly coincide for the same system parameters.

If at the initial time moment concurrence is not equal to zero
[$C(0)\neq0$] there can exist a time moment $t=t_0$, when
concurrence turns to zero [$C(t_0)=0$] (see Fig.\ref{figure2}).

\begin{eqnarray}
t_{0}=\frac{1}{2( \lambda_{S^{0}}-\lambda_{T^{0}})}\cdot
ln(\frac{N_{S^{0}}}{N_{T^{0}}}).
\end{eqnarray}

Further system time evolution leads to the concurrence increasing
reaching its maximum value

\begin{eqnarray}
C(t_{max})&=&N_{T^{0}}\cdot
e^{\frac{-\lambda_{T^{0}}}{\lambda_{S^{0}}-\lambda_{T^{0}}}\cdot
ln(\frac{N_{S^{0}}\cdot \lambda_{S^{0}}}{N_{T^{0}}\cdot
\lambda_{T^{0}}})}-\nonumber\\&-&N_{S^{0}}\cdot
e^{\frac{-\lambda_{S^{0}}}{\lambda_{S^{0}}-\lambda_{T^{0}}}\cdot
ln(\frac{N_{S^{0}}\cdot \lambda_{S^{0}}}{N_{T^{0}}\cdot
\lambda_{T^{0}}})}
\end{eqnarray}

at time moment

\begin{eqnarray}
t_{max}=t_{0}+\frac{1}{2( \lambda_{S^{0}}-\lambda_{T^{0}})}\cdot
ln(\frac{\lambda_{S^{0}}}{\lambda_{T^{0}}}).
\end{eqnarray}

So, concurrence could reveal non-monotonic behavior for mixed
two-electronic initial state with opposite spins. We would like to
mention that despite the fact that both theoretical approaches give
the same result for the considered system of two coupled quantum
dots with Coulomb correlations there is some difference between
them. Theoretical approach based on the equations of motion for
localized electrons occupation numbers (see Section \textbf{A})
provides possibility for the analysis of concurrence and spin
correlations time dependent behavior in the complicated systems of
many correlated coupled quantum dots taking into account all orders
localized electron correlation functions and considering
interference effects. A closed system of equations can be obtained
for an arbitrary number of quantum dots in the situation of weak
coupling to reservoir, but these equations will have a rather
cumbersome form and can be hardly solved analytically. Theoretical
approach based on the pseudo-particles formalism (see Section
\textbf{C}) is more straightforward and provides the possibility to
analyze analytically time evolution of the degree of entanglement in
the system of quantum dots with a small number of available
electronic states in the case of weak interaction between correlated
quantum dots and reservoir. To conclude, both methods can be applied
for the localized charge and spin kinetics analysis in coupled
quantum dots with Coulomb correlations, but for the complicated
systems method based on the equations of motion for localized
electrons occupation numbers is more preferable. However, systems
with a small number of available electronic states could be better
analyzed by means of pseudo-particles formalism as it provides the
possibility to obtain explicit expressions for the time evolution of
system characteristics.

\section{Conclusion}

Time evolution of initially prepared entangled state in the system
of correlated coupled quantum dots has been analyzed by means of two
different approaches. The first one is based on the equations of
motion for all orders localized electron correlation functions
taking into account interference effects. The second approach deals
with the kinetic equations for pseudo-particle occupation numbers
considering constraint on the possible physical states. Both
approaches allow us to follow the changes of the entanglement during
time evolution of the two coupled quantum dots system due to the
concurrence direct link with quantum dots pair correlation
functions. For different initial mixed states the concurrence
(degree of entanglement) could reveal non-monotonic behavior and
even considerably increase during the time evolution of quantum dots
system. Obtained results reveal the possibility of system switching
between entangled and unentangled (concurrence is equal to zero)
states during the relaxation process. This fact provides the method
of controllable tuning of the degree of entanglement for the
electronic quantum dots systems based on the analysis of its
non-stationary characteristics.

This work was supported by Russian Science Foundation (project no.
$16-12-00072$). V.N.M. also acknowledge the support by the RFBR
grant $16-32-60024$ $mol-a-dk$.

 \pagebreak

\end{document}